\documentclass[12pt,preprint]{aastex}

\newcommand{\goes}{\textit{GOES}}
\newcommand{\ha}{H$\alpha$}
\newcommand{\trace}{\textit{TRACE}}
\newcommand{\sm}{$\sim$}
\newcommand{\Sdo}{\textit{Solar Dynamics Observatory}}

\defcitealias{masson09}{M09}
\defcitealias{pariat10}{P10}

\begin{document}

\title{CIRCULAR RIBBON FLARES AND HOMOLOGOUS JETS}
\author{Haimin Wang and Chang Liu}
\affil{Space Weather Research Laboratory, Center for Solar-Terrestrial Research,\\New Jersey Institute of Technology, University Heights, Newark, NJ 07102-1982, USA}

\email{haimin.wang@njit.edu}

\begin{abstract}
Solar flare emissions in the chromosphere often appear as elongated ribbons on both sides of the magnetic polarity inversion line (PIL), which has been regarded as evidence of a typical configuration of magnetic reconnection. However, flares having a circular ribbon have rarely been reported, although it is expected in the fan--spine magnetic topology involving reconnection at a three-dimensional (3D) coronal null point. We present five circular ribbon flares with associated surges, using high-resolution and high-cadence \ha\ blue wing observations obtained from the recently digitized films of Big Bear Solar Observatory. In all the events, a central parasitic magnetic field is encompassed by the opposite polarity, forming a circular PIL traced by filament material. Consequently, a flare kernel at the center is surrounded by a circular flare ribbon. The four homologous jet-related flares on 1991 March 17 and 18 are of particular interest, as (1) the circular ribbons brighten sequentially, with co-spatial surges, rather than simultaneously, (2) the central flare kernels show an intriguing ``round-trip'' motion and become elongated, and (3) remote brightenings occur at a region with the same magnetic polarity as the central parasitic field and are co-temporal with a separate phase of flare emissions. In another flare on 1991 February 25, the circular flare emission and surge activity occur successively, and the event could be associated with magnetic flux cancellation across the circular PIL. We discuss the implications of these observations combining circular flare ribbons, homologous jets, and remote brightenings for understanding the dynamics of 3D magnetic restructuring.

\end{abstract}

\keywords{Sun: activity  -- Sun: flares -- Sun: magnetic topology}

\section{INTRODUCTION} \label{introduction}
The elongated structures of solar flare emissions, as the name ``flare ribbons'' implies, have long been observed at \ha\ and UV wavelengths \citep[e.g.,][]{zirin88}. A pair of flare ribbons residing in opposite magnetic polarities run parallel to the magnetic polarity inversion line (PIL) lying between them, and the ribbons separate from each other in the direction perpendicular to the PIL. Such a configuration and its kinematics has been regarded as evidence of the classical two-dimensional-like (2D) magnetic reconnection model called the CSHKP model \citep{carmichael64,sturrock66,hirayama74,kopp76}, in which two ribbons form as the energy release from a series of coronal X-points along an arcade of loops produces bright flare emissions at the loop footpoints, and the ribbon separation is resulted from the successive reconnections of higher coronal arcades.

Nevertheless, actual flares take place in a more complicated three-dimensional (3D) structure, hence the standard model, despite of its general applicability \citep{hudson11}, may not explain some features of the flare ribbon evolution, such as the frequently observed expansion of ribbons along the PIL before the perpendicular separation \citep[e.g.,][]{fletcher04}. In relation to magnetic topology, chromospheric flare ribbons are generally situated at locations intersected by separatrices dividing domains of distinct connectivity \citep[e.g.,][]{mandrini91}, or quasi-separatrix layers (QSLs) possessing strong connectivity gradients \citep[e.g.,][]{demoulin97}. This is because that intense current sheets are preferentially built up at separatrices/QSLs \citep[e.g.,][]{lau90,aulanier05}, along which reconnection-accelerated particles can precipitate into the lower atmosphere. It is known that separatrices can be typically generated by magnetic null points, which are common structures in the corona due to the mixed-polarity nature of the photospheric fields \citep{schrijver02}. Magnetic field lines associated with a 3D coronal null point usually display a fan-spine configuration \citep{lau90,torok09}, where the dome-shaped fan portrays the closed separatrix surface and the inner and outer spine field lines in different connectivity domains pass through the null point. The footpoint of the inner spine has a magnetic polarity opposite to those of the fan, which forms a circular PIL. Magnetic reconnection can be induced in such single null points, as the fan/spines deviate from the null when subject to shearing or rotational perturbations \citep[e.g.,][]{pontin07a,pontin07b}. It is then expected that as a result of null-point reconnection, flare emissions at the footpoints of the fan field lines would constitute a closed circular flare ribbon, and that the spine-related flare footpoint would be a compact source. Surprisingly, there seems to be very few events reported in the literature that assume the shape of circular ribbons in \ha\ \citep{sundara94} or UV \citep{ugarte07,su09} wavelengths.

In retrospect, the most comprehensive study of circular ribbon flares was carried out by \citet[][hereafter M09]{masson09} and \citet{reid12} for a confined C8.6 flare on 2002 November 16 using 1600~\AA\ UV continuum images from \trace. Several observational features are prominent in this event. First, three ribbons were present, including two elongated ribbons inside and outside the circular ribbon, respectively. Second, the circular ribbon brightened sequentially in the counterclockwise direction. Third, the appearance of the outside, remote ribbon had a \sm30~s delay relative to the main ribbons. A potential field extrapolation revealed that these ribbons indeed seem to map the photospheric intersections of the fan and spine field lines, which stem from a coronal null point (see Figure~\ref{f1}). With the aid of 3D MHD numerical simulations, the authors further suggested that the fan and spine separatrices are embedded in larger QSLs. The extended shape of QSLs surrounding the singular spine field lines is consistent with the observed elongated spine ribbons. Importantly, field lines can undergo slipping and slip-running reconnection within the QSLs \citep{aulanier06}. Their results that field lines closer to the null would reconnect first and that the slipping motion is toward the null could then account for the counterclockwise propagation of the circular ribbon emission. The implied sequential occurrence of slipping/slip-running reconnection within the fan and null-point reconnection involving the outer spine may explain the delayed brightening of the remote ribbon corresponding to a second phase of flare emission.

By flipping the outer spine field lines in \citetalias{masson09} to open outward, one could arrive at an axisymmetrical null-point and fan-spine topology (see Figure~\ref{f1}), which \citet{pariat09} used to model solar polar jets. Jets in the solar atmosphere can appear in a variety of forms that may be interrelated \citep[e.g.,][]{chae99}, such as the cool ejections (surges) in \ha\ \citep[e.g.,][]{schmieder84} and the hot ejections (jets) in EUV/UV \citep[e.g.,][]{alex99} and soft X-rays (SXRs) \citep[e.g.,][]{shibata92}. In fact, the above configuration can be simply developed by flux emergence into a unipolar region such as the polar coronal holes, a scenario typically assumed for jets \citep[e.g.,][]{shibata92}. In the model of \citet{pariat09}, the imposed twisting motion within the fan circle at the photospheric boundary builds up magnetic stress, until an ideal instability sets in to cause interchange reconnection at the null, driving massive, high-speed jets. The numerical investigation was then extended in \citet[][hereafter P10]{pariat10} by tilting the outer spine to break the axisymmetry and applying a constant stress to both closed- and open-connectivity domains (see Figure~\ref{f1}). Under these prescribed perturbations, it was then shown that 3D null-point topology can naturally produce successive, homologous jets, which are frequently observed \citep[e.g.,][]{chifor08b}. The axes of the modeled homologous jets rotate in the same direction that is opposite to the boundary driving direction. The simulations also demonstrated that the twist originally stored in the closed domain (e.g., emerging fluxes) can be transferred to the reconnected, open field lines, which explains the observed unwinding motion and helical structure in jets \citep[e.g.,][]{canfield96,liu11,liu+wei11}. Moreover, the authors pointed out that even though there should form a circular ribbon related to the fan and another ribbon corresponding to the inner spine, the properties of flare ribbons on the surface may be hard to predict, due to the dynamic nature of the 3D reconnection at the null point. To our knowledge, jet-associated circular ribbon flares have not been reported before.

In this paper, we present five flares including four jet-related events, the ribbons of which all have a circular shape. It is interesting that three homologous flares are also accompanied by a remote ribbon. Our events thus combine the key observational features on which the models of \citetalias{masson09} and \citetalias{pariat10} are based, incorporating circular flare ribbons, homologous jets, and remote brightenings. The uncovering of these observations benefits from studying a sample portion of the historical films of the Big Bear Solar Observatory (BBSO), which was established and previously directed by Professor Hal Zirin. The plan of this paper is as follows: in Section~\ref{data}, we describe the data sets and reduction procedure. In Section~\ref{results}, we present the main results of data analysis and discuss their implications. Major findings are summarized in Section~\ref{summary}.

\section{OBSERVATIONS AND DATA REDUCTION} \label{data}

From 1969 to 1995, BBSO recorded all observations on 35~mm films before its post-1995 switch to digital imaging. The film data include full-disk \ha\ observations from its 8~inch telescope and high-resolution observations in H$\alpha$ and He~{\sc i} D3 from its 10 and 26~inch telescopes. The cadence of data ranges from \sm10~s to 1~minute. Although the data were used by BBSO researchers and visitors who published many papers, their scientific capability has far from been fully explored, mainly because of the difficulty in accessing and evaluating films and the lack of data reduction (e.g., image alignment). With the BBSO/NJIT digitization project \citep{liu10b}, we have digitized all the full-disk and selected high-resolution film images. The digitization is carried out using a commercial film digitizer manufactured by Walde Inc., which converts film images to 12~bit digital data with a 2048~$\times$~1600 pixels resolution. The final image product is stored in the standard FITS format, and has a pixel scale of \sm0.15\arcsec\ to 1\arcsec. We understand that the digitized images may not be suitable for accurate photometric analysis due to the nonlinear intensity response of films. Nevertheless, morphological study on solar features (e.g., flares, filament activities, and surges) can be carried out after applying standard image processing techniques.

The events in the present study are selected from our survey on activities in 1991 during the peak of solar cycle 22. We mainly use the high-resolution (\sm0.15 pixel$^{-1}$) and high-cadence (\sm15~s) \ha\ images with a field of view (FOV) of \sm300\arcsec~$\times$~220\arcsec\ obtained with the 10~inch telescope. The observations were mainly made in \ha~$-$~0.6~\AA, with occasional scanning from \ha~$-$~2~\AA\ to \ha~$+$~2~\AA\ in a step size of 0.25~\AA. For optimal contrast, the images were transformed to 8~bit using a nonlinear conversion table supplied by the digitizer. The image alignment was implemented with sub-pixel precision, and intensity was normalized to that outside the flaring region in a quiet-Sun area. A de-stretching algorithm \citep{shine94} using running references was applied in order to further reduce the atmospheric distortion.

The full-disk line-of-sight (LOS) magnetic field structure of the photosphere was observed with the 512-channel Magnetograph \citep{livingston76} in Fe~{\sc i}~8688~\AA\ using the vacuum telescope of the National Solar Observatory (NSO) at Kitt Peak (KP). We registered \ha\ images with NSO/KP magnetograms by matching sunspot and plage areas, with an alignment accuracy within about 5\arcsec. For the 1991 February 25 event, we also used magnetic field data taken by BBSO's Videomagnetograph system \citep[VMG;][]{mosher77}, which has a spatial resolution around 1\arcsec\ and cadence around 1 minute. The data were already in the digital form and were calibrated using NSO/KP magnetograms as reference. To show the temporal evolution of the flares and understand the relationship between the main flares and remote brightenings, we used SXR time profiles from \goes~7 satellite. The time derivative of the SXR flux represents a proxy for the hard X-ray light curve \citep{neupert68}.

\section{RESULTS AND ANALYSIS} \label{results}
In this section, we describe the major observational features found in the digitized BBSO film data set of five events as listed in Table~\ref{table1}. In particular, the events 1--4 during 1991 March 17--18 represent homologous circular ribbon flares and surges. We also discuss the results under the context of the models of \citetalias{masson09} and \citetalias{pariat10} for the 3D reconnection in a null-point topology, whenever a plausible comparison can be made. More dynamical detail can be seen in the accompanying mpeg animations\footnote{The images in the animations were not corrected for the solar p-angle.}

\subsection{Event 1: Circular Flare Ribbon and Surge} \label{circular}

\begin{figure}
\epsscale{1}
\plotone{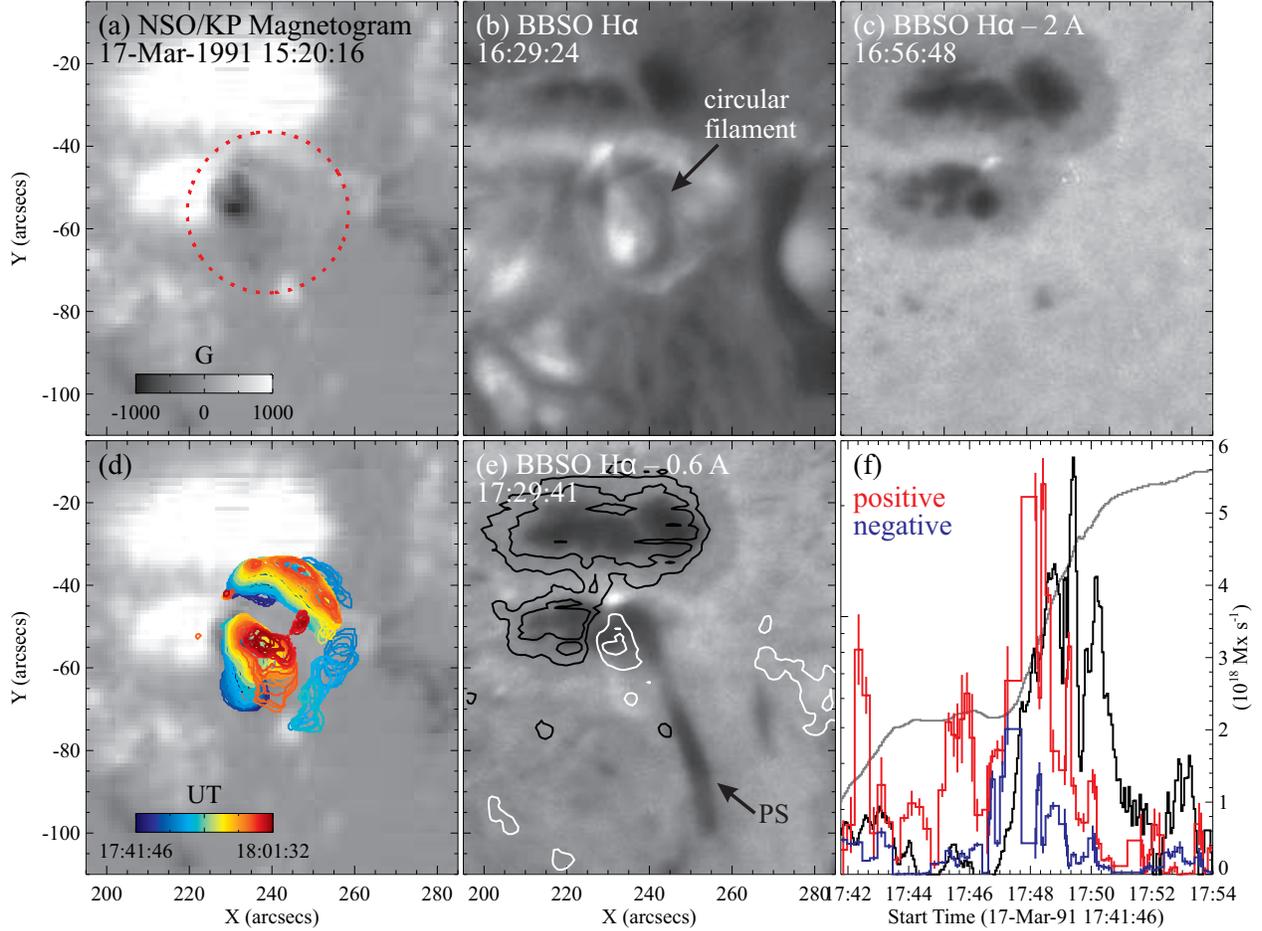}
\caption{The source active region NOAA 6545 for the events 1--4 in Table~\ref{table1}. A $\delta$ magnetic configuration is demonstrated by comparing the magnetic field (a) and sunspot structure seen in the far blue wing of \ha\ (c). The parasitic, negative field is surrounded by positive field (as illustrated using the red dotted line in a), forming a circular PIL that is also traced by a filament (b). Pre-flare surge (PS) activity seems to originate from the northern side of the parasitic field (e). The levels of magnetic field contours in (e) are $-$600, $-$200, 600, and 1000~G. Time profiles of the positive/negative (red/blue) flux change rates of the event 1 (f; see Figure~\ref{f3}) are derived by tracking the ribbon motion (color-coded contours in d), and are compared with those of \goes\ 1--8~\AA\ flux (gray) and its time derivative (black). \label{f2}}
\end{figure}

The active region of interest NOAA AR 6545 emerged on the east limb on 1991 March 10. It appears in the $\beta\gamma\delta$ magnetic configuration till March 16 when it rotated to the disk center. The region is the source region for five X-class and nine M-class flares. According to the USAF/NOAA report, the region showed slight decay and possible magnetic simplification during March 17--18, while there is still clear evidence of a $\delta$ configuration in the bottom part of the leading spot (cf. Figures~\ref{f2}(a) and (c)). As an obvious property of magnetic field structure of this region, the central negative field seems to be encircled by the surrounding positive field, which forms a quasi-circular PIL (Figure~\ref{f2}(a)). The circular shape of the source region and a filament lying along the PIL are also clearly visible in \ha\ line center (Figure~\ref{f2}(b)).

\begin{figure}
\epsscale{1.0}
\plotone{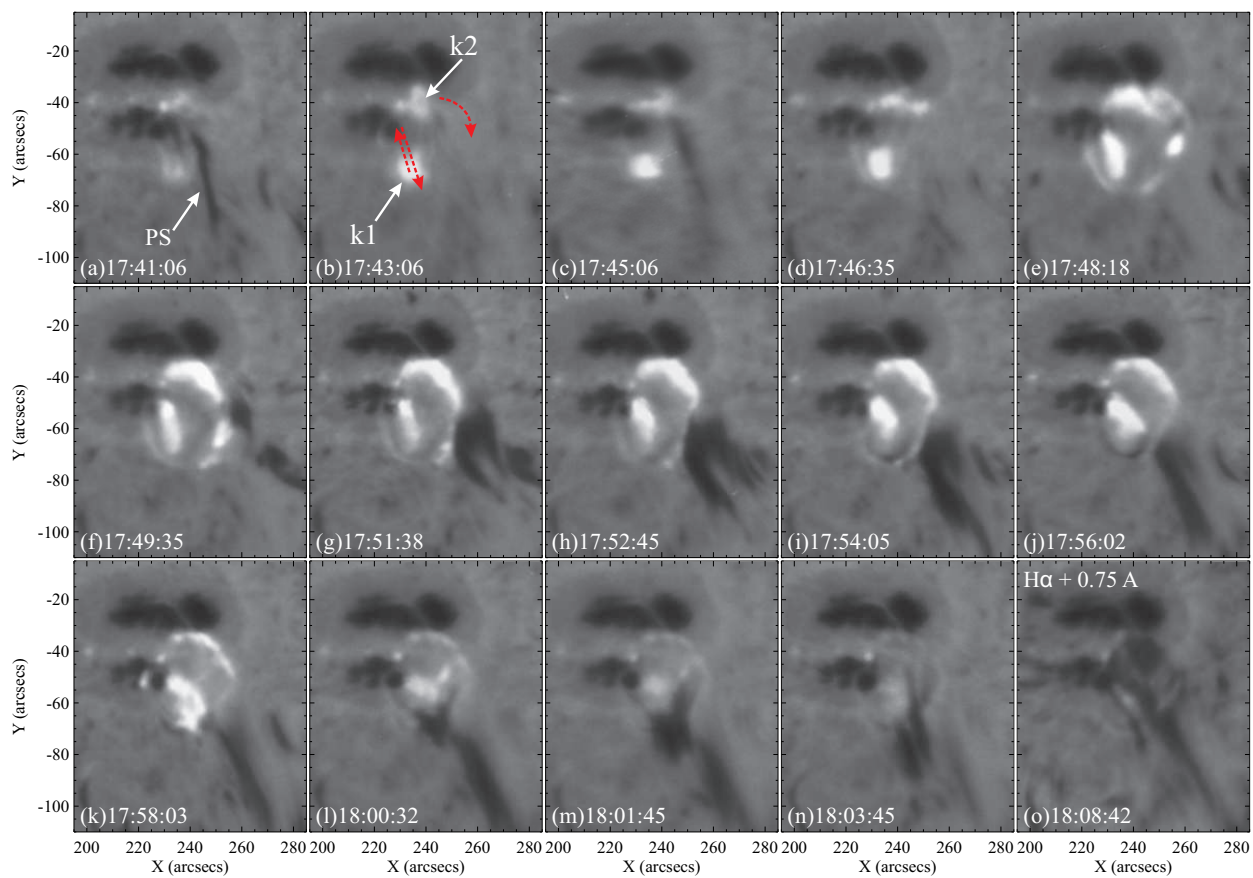}
\caption{Time sequence of H$\alpha - 0.6$~\AA\ images of the event 1. The central ribbon k1 first moves northeastward then southwestward. The outer circular ribbon k2 brightens first in the north (first row), and then propagates clockwise toward south with associated surges (c--k). Another episode of ejection occurs possibly due to the partial eruption of the circular filament (l-n). These motions are illustrated using red arrows. PS denotes the pre-flare surge as in Figure~\ref{f2}. All the images are aligned with respect to 1991 March 17 17:25:56 UT. \label{f3}}
\end{figure}

From \sm17:10~UT on March 17 before the event occurrence, a dark preflare surge (PS) apparently rooted at the northern side of the central negative field gradually extends upward to form a long and slim structure (see Figure~\ref{f2}(e)). This outward ejection as observed in \ha\ blue wing seems to subside after \sm17:32~UT, while it picks up again from \sm17:37~UT till the event onset around 17:42~UT (Figure~\ref{f3}(a)). The M2 flare starts with rapid brightenings of the kernels k1 and k2 located in the negative and positive field regions, respectively (cf. Figures~\ref{f3}(b) and \ref{f2}(a)). As the flare progresses, we note the following dynamic activities. First, k2 shows sequential brightening in the clockwise direction, moving from the top to the bottom of the circular region from \sm17:45 to 17:51~UT with an average speed of \sm100~km~s$^{-1}$. This motion also seems to be apparently accompanied by the dark surge ejection (Figures~\ref{f3}(c)--(k)). Weaker ribbon brightenings in the south are discernible from \sm17:48~UT, so that the entire outer ribbon appears as a circle (Figure~\ref{f3}(e)). The circular shape of the flare ribbon and the progressive brightening along the rim of the circle are directly reminiscent of the event of \citetalias{masson09}. Meanwhile, the clockwise drifting of the surge axis seems in line with the model of \citetalias{pariat10}. Second, k1 first moves northeastward from \sm17:46~UT at \sm40~km~s$^{-1}$, and after k1 reaches the strongest negative field region at \sm17:49~UT (Figure~\ref{f3}(f)), intense surge associated with the circular ribbon in the west begins to eject outward. Subsequently, k1 turns back to propagate southwestward at \sm25~km~s$^{-1}$ during \sm17:52--17:58~UT. There is a separate episode of ejection possibly due to the partial eruption of the circular filament (Figures~\ref{f3}(l)--(n)). k1 thus evolves from compact to elongated shape and undertakes an intriguing ``round-trip'' motion (illustrated in Figure~\ref{f3}(b)).

To track the flare ribbon quantitatively, we use the intensity-based binary masks method \citep[e.g.,][]{liu07b,liu09b} and depict the aforementioned motion of k1 and k2 from \sm17:42 to 18:01 as color-coded areas over-plotted on the co-aligned NSO/KP magnetogram in Figure~\ref{f2}(d). In the standard 2D model, the flare ribbon motion is a direct mapping of energy release via coronal magnetic reconnection, the rate of which can be evaluated using the magnetic flux change rate of the chromospheric ribbon area as $\dot\phi = (\partial / \partial t) \int B\ da$, where $B$ is the vertical magnetic field component of the surface ribbon element $da$. For this event near the disk center, we take the LOS magnetic field measurement and in order to estimate uncertainties of the ribbon selection, we vary the intensity cutoff value (about 160) within $\pm$10\% to obtain the error bars. The resulting magnetic flux change rate in both the positive ($\dot\phi_+$) and negative ($\dot\phi_-$) fields as a function of time are plotted in Figure~\ref{f2}(f). It can be seen that $\dot\phi_+$ and $\dot\phi_-$ are temporally correlated with the flare non-thermal emission (approximated using the time derivative of the SXR light curve), which is theoretically expected \citep{priest02} and may demonstrate the essential role of magnetic reconnection in producing this event. We note that the intensity threshold used in selecting ribbons does not affect the overall timing (e.g., the peak time) of the derived $\dot\phi$ \citep{liu07b}. However, the results of $\dot\phi_{+(-)}$ also obviously show that the magnetic flux change in two polarities are grossly unbalanced \citep[cf.][]{qiu05,liu07b}. We are cautious that (1) the calculation may be affected by the limitation of the quality of the digitized film data (such as saturation seen at the center of flare ribbons), (2) the measurement does not take into account the reconnected flux associated with any open fields, and (3) most importantly, our method is only applicable if the event configuration can be approximated by a 2D-like picture.

On the other hand, it is noteworthy that the event exhibits no clear evidence of the eruption of the PIL filament (Figure~\ref{f2}(b)) preceding the event initiation, while possesses many properties (e.g., circular ribbon and drifting surge) that are very similar to the observations and models of \citetalias{masson09} and \citetalias{pariat10}. This strongly implies that a 3D reconnection possibly associated with a coronal null point, rather than the standard 2D scenario, could be the flare mechanism of this event. We make such considerations further in the following.

First, the presence of a coronal null point in this event could already be well indicated by the circular shape of the outer flare ribbon k2 \citepalias{masson09}. The fan-spine magnetic topology may be envisioned by the result of potential field extrapolation performed using the NSO/KP magnetogram at March 17 15:20~UT (see Figure~\ref{f4}(a)), in which red (resp. blue) lines denote closed (resp. open) fields. The expected dome-shaped magnetic structure is also discernible by the black surges observed in Figure~\ref{f3}(o). Similar to \citetalias{masson09}, we thus link the circular ribbon k2 and the inner ribbon k1 to the chromospheric mapping of the fan and the inner spine field lines (also cf. Figure~\ref{f1}), respectively. Another notable observation is the slim surge right before this event (Figures~\ref{f2}(e) and \ref{f3}(a)). Its root is close to the northeast corner of the PIL, and the entire surge seems to lie directly above the circular source region (cf. Figures~\ref{f2}(b) and (e)). These are obviously distinct from the later surge that are co-spatial with the circular ribbon (Figures~\ref{f3}(c)--(k)). We speculate that (1) this slim surge could be the precursor of this event tracing the outer spine field lines, and that (2) the coronal null point might be near the northeast resulting in a non-axisymmetric fan-spine magnetic configuration. The latter may be implied by the asymmetric surface field distribution within the circle (Figure~\ref{f2}(a)), and has been shown to be a favorable condition for the 3D reconnection \citepalias{masson09,pariat10}.

Second, the clearly observed sequential brightening of the fan ribbon k2 and the elongated shape of the inner spine ribbon k1 well resemble the event of \citetalias{masson09}, and would strongly suggest the existence of extended QSLs, in which the null point and the related fan-spine structure are embedded. In the model of \citetalias{masson09}, field lines within the QSLs first undergo slipping and slip-running reconnection toward the null point before the occurrence of the null-point reconnection. In a second phase, field lines encounter slip-running and slipping reconnection as they slip away from the null. These modeled results agree with the ordered motion of the fan and spine ribbons in their event observed in \trace\ 1600~\AA. We note that such UV emission has a slow decay due to slow cooling or gradual heating not related to the non-thermal electrons \citep{qiu10}. In contrast, blue wing \ha\ images may better represent the precipitation of energetic electrons in the lower atmosphere \citep[e.g.,][]{lee06} hence may better trace the motion of flare ribbons/kernels. Besides the clearly observed sequential brightening of k2, a prominent finding of our observations is thus that the inner spine ribbon/kernel k1 first moves northeast presumably toward the null point and then reverts back. These observations may then well reflect the different phases of the QSL reconnection involving the succession of the slipping and slip-running reconnection. It is pertinent to point out that they are not the same field lines that slip reconnect through the QSL before and after the null-point reconnection \citepalias{masson09}. Moreover, the intense surge, which should be resulted from the null-point reconnection involving open outer spine field line \citepalias{pariat10}, occurs after k1 reaches the northern turning point. This could be consistent with the view that the eruptive null-point reconnection can only be triggered when the system has stored a sufficient amount of energy \citepalias{masson09}. The measured average velocities of k1 and k2 are comparable with or larger than the estimated slippage velocity of 30~km~s$^{-1}$ in the model of \citetalias{masson09}. It is also worthwhile to mention that different from the confined event of \citetalias{masson09} where the null-point reconnection is related to remote brightenings, our eruptive event exhibits jet activities as the signature of null-point reconnection as in \citetalias{pariat10}.

\subsection{Events 2--4: Circular Flare Ribbon, Surge, and Remote Brightening} \label{remote}
This $\delta$ region in a circular shape produced other three flares in about one day, including an X1 (event 2), an M5 (event 3) and an M1 (event 4) flares. These three events show similar characteristics in the following aspects: (1) they all have a circular ribbon that encompasses another smaller elongated ribbon, and the ribbons evolve in nearly the identical regions. (2) In all the events, the circular ribbon brightens sequentially in a clockwise fashion, in tandem with the co-spatial surge (except the event 4); meanwhile, the inner ribbon shows a similar ``round-trip'' motion as elaborated in Section~\ref{circular}. It is however noticeable that these ribbon dynamics are most pronouncedly observed in the event 1 with the longest flare duration (see Table~\ref{table1}). These properties thus connote that the events 1--4 constitute homologous flares and surges (whilst the event 4 shows insignificant surge activity probably consistent with its lowest flare magnitude), and that they could be associated with similar magnetic reconnection processes.

\begin{figure}
\epsscale{1.0}
\plotone{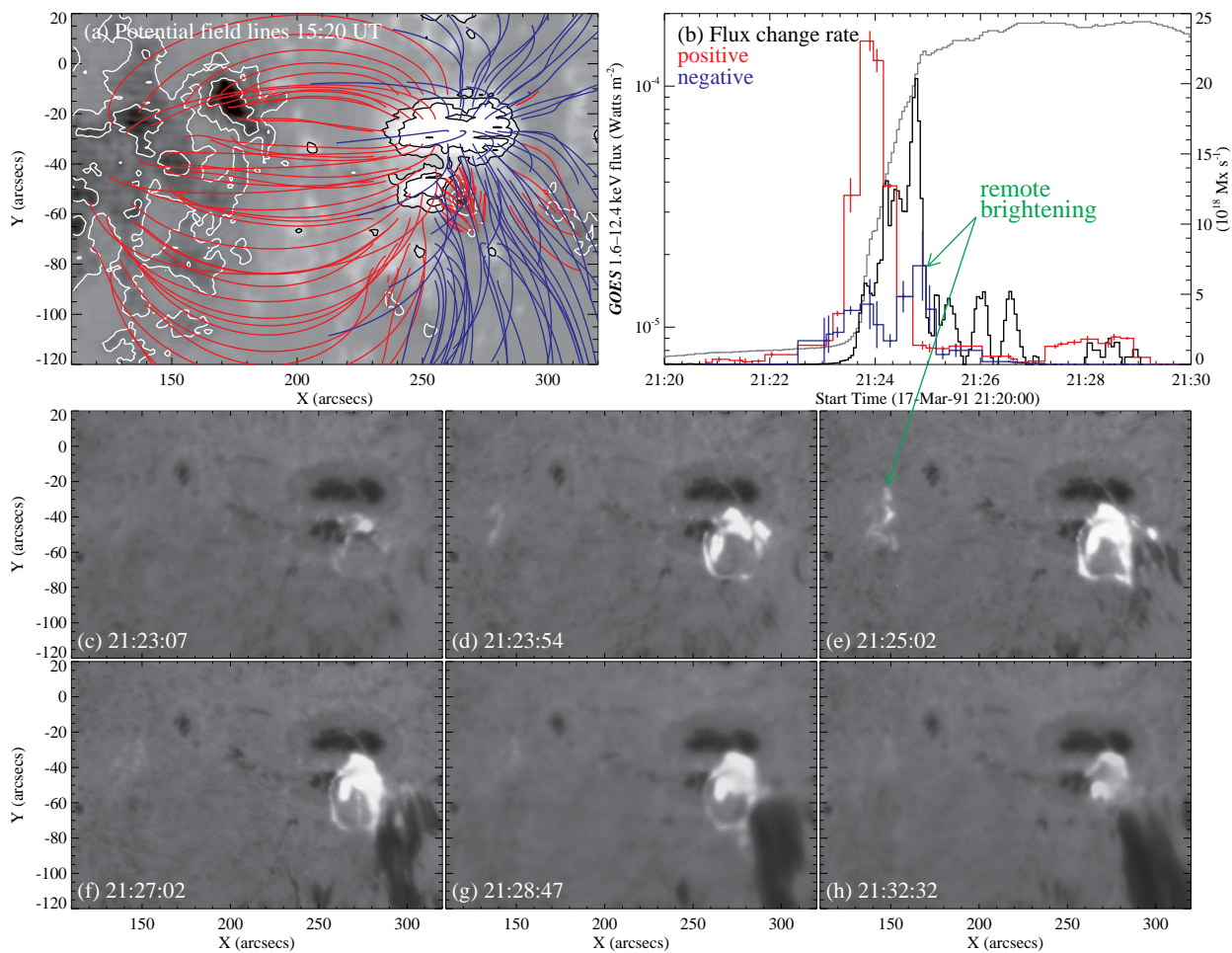}
\caption{Magnetic field structure and event evolution of the event 2. (a) A NSO/KP magnetogram over-plotted with the extrapolated potential field lines. Red and blue lines indicate closed and open fields, respectively. The levels of magnetic field contours are the same as those in Figure~\ref{f2}(e). (b) Same as Figure~\ref{f2}(f) but for this event. The peaks around 21:25~UT correspond to the brightening of the remote ribbon. (c--h) Time sequence of \ha~$-0.6$~\AA\ images. All the images are aligned with respect to 1991 March 17 21:13:02~UT. \label{f4}}
\end{figure}

\begin{figure}
\epsscale{1.0}
\plotone{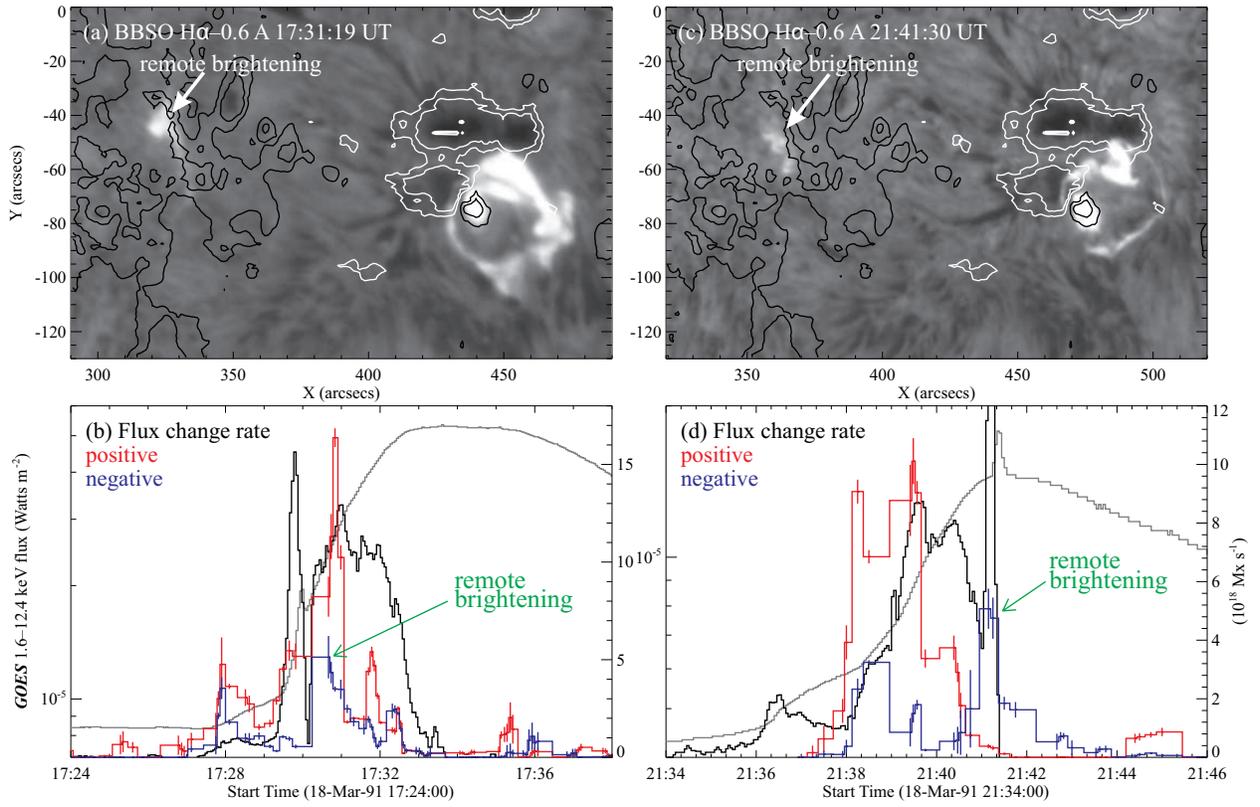}
\caption{Results of the events 3 (a--b) and 4 (c--d). \ha\ images showing the main flare region (including a circular ribbon and an inner small ribbon) and a remote ribbon are over-plotted with contours of magnetic field, with the same levels as in Figure~\ref{f2}(e). \label{f5}}
\end{figure}

Nevertheless, different from the event 1, a conspicuous common feature of the events 2--4 is the existence of a third elongated flare ribbon brightening in the eastern plage region of negative polarity (same as the central parasitic field), which is about 120\arcsec\ apart from the main circular flaring site (see Figures~\ref{f4}(c)--(h), \ref{f5}(a)(c)). Remote brightenings are often seen in large eruptive flares, and have been understood as the result of the direct heating by hot electrons at relativistic speed traveling along large-scale, closed magnetic loops connecting the main flare to remote sites \citep[e.g.,][and references therein]{liu06}. This is most likely the mechanism for our event, in which the remote flare ribbon brightened almost simultaneously, with a minor trend of drifting motion toward the south. It is unlikely that these brightening patches are the sequential chromospheric brightenings observed to propagate away from the flare site \citep{balasubramaniam05}. The present remote ribbons also do not bear resemblance to those described in \citet{wang05}, where the remote brightenings were interpreted as the interaction between the erupting fields with the nearby larger scale fields. In view of the analogous study of \citetalias{masson09} and \citetalias{pariat10}, it is therefore appealing to conjecture that some outer spine fields dip down to the remote brightening region, while some open upward related to the surges, and that both the remote brightening and the surge are due to the null-point reconnection involving the outer spine field lines. The corresponding magnetic structure as sketched in Figure~\ref{f1} based on our observations, which can also be envisioned from the extrapolated fields (Figure~\ref{f4}(a)), is hence a combination of those depicted in \citetalias{masson09} and \citetalias{pariat10}.

Another persuasive argument for the close linkage between the remote flare ribbon and the main circular flare lies in the fact that the remote brightenings in the events 2--4, as manifested by the peaks of $\dot\phi_{-}$, are all co-temporal with a second phase of flare energy release signaled by a separate peak in the time derivative of the SXR flux. This is clearly evidenced in Figures~\ref{f4}(b) and \ref{f5}(b)(d). We are convinced that the peaks of $\dot\phi_{-}$ are caused by the remote brightenings, as those peaks would disappear if only the inner spine ribbons in the negative field region are included. We also note that there is a \sm1~minute lag between the peak of the remote brightenings and the main energy release of flares, the latter of which is reflected by the earlier peaks in the SXR flux derivative and $\dot\phi$. As implied by \citetalias{masson09} and elaborated in \citet{reid12}, we tend to interpret this time delay as the time related to the slipping of the magnetic fields within the QSLs toward the null point. Other possibilities for the delay include that (1) it is the time required to store enough energy in the system to trigger an impulsive surge ejection, and (2) since the outer spine could be initially open that is noticeable from the magnetic field extrapolation of the active region (Figure~\ref{f4}(a)), the delay might be due to the time required for the outer spine to close down to the photosphere to produce remote brightenings. A more intricate issue is posed by the presence of both the remote brightenings and surges in the events 2--3, in which the peak of the remote brightenings appears to coincide with the start of surges as seen in the time-lapse movies. We surmise that this may indicate a change of magnetic topology quasi-instantaneously. Specifically, our observations could be the first ones that can be interpreted as the signature of the opening and closing of the outer spine in a null-point topology.

\subsection{Event 5: Circular Flare Ribbon, Surge, and Flux Cancellation}
\begin{figure}
\epsscale{1.0}
\plotone{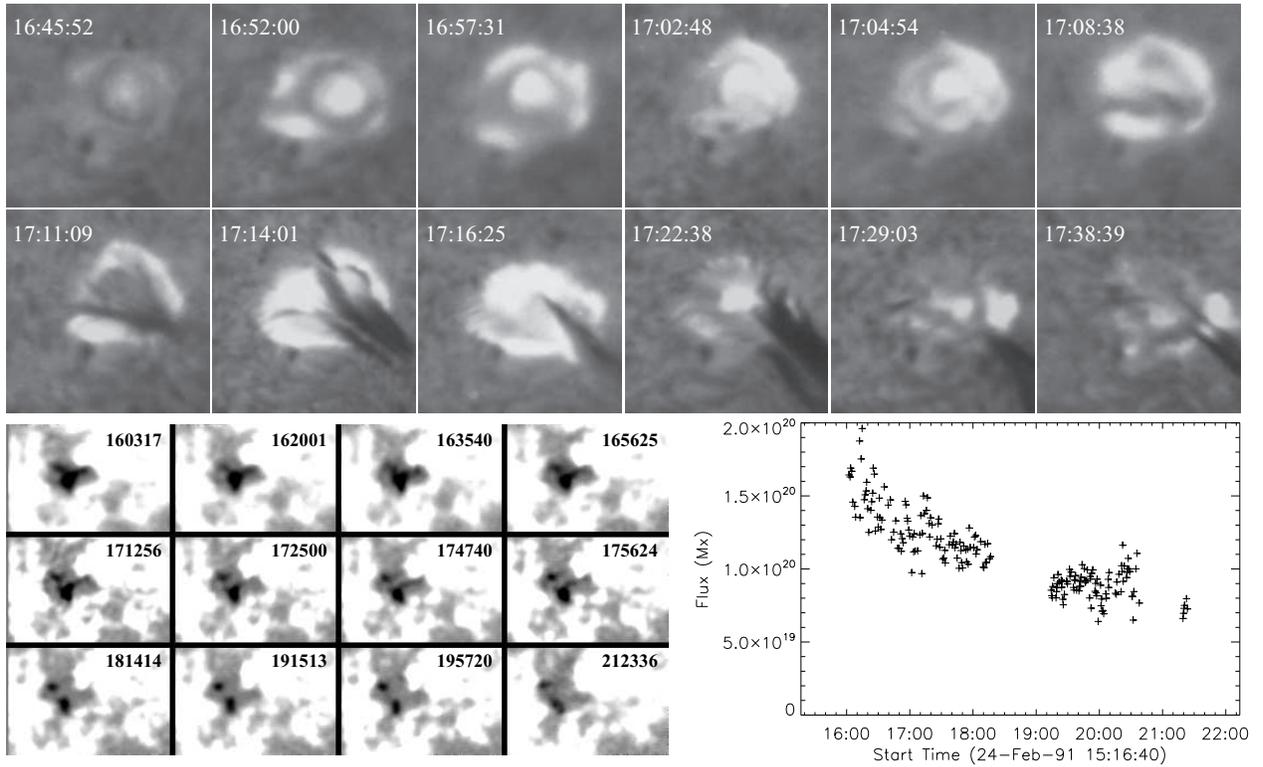}
\caption{Upper panels: Time sequence of H$\alpha - 0.6$~\AA\ images of the event 5. The FOV is 46\arcsec\ $\times$ 46\arcsec, at 26$^{\circ}$ measured clockwise from the solar north. Lower left panels: Time series of BBSO VMG magnetograms (scaled from -200 to 200~G) around the flaring region, highlighting the circular PIL. Lower right panel: Time profile of the total negative magnetic flux, demonstrating the magnetic flux cancellation across the circular PIL. \label{f6}}
\end{figure}

This flare is associated with the parasitic negative field in the leading positive plage region of NOAA AR 6509 on 1991 February 25 (NSO/KP magnetogram was not available for this day). The findings of the filament material lying along the circular PIL and the overall sequential brightening of the fan-related circular flare ribbon, as shown in the Figure~\ref{f6} (upper panels) and the movie, are similar to those of the events 1--4. It is nevertheless obvious that the inner spine ribbon remains compact throughout the event period. According to \ha\ images, this circular ribbon appears more circular than those in the other events, which suggests that the fan-spine topology is more symmetric in this event. Therefore, the QSL around the inner spine is expected to be less squashed, which implies that the inner spine ribbon should show a more compact form. Another notable feature of this event is the well separated phases of the circular ribbon flare (from \sm16:49--17:10~UT) and the surge (after \sm17:10~UT). This may provide another example in which the slipping/slip-running reconnection and the null-point reconnection may occur successively in a single event \citepalias{masson09}. As a side note, an additional brightening is discernible from \sm17:17:29 (see the animation) during the phase of surge in a region close to the main sunspots. However, the timing of the brightening seems not to be related to significant peaks in X-rays (cf. the events 2--4).

We use the available BBSO VMG observations to study the associated magnetic field evolution. From Figure~\ref{f6} (lower left panels), one can see that the central negative field is broken into several fragments, which move toward the PIL and cancel with the nearby positive field. Since the positive flux covers a much larger area, it is hard to calculate the small amount of change due to flux cancellation. Instead, we plot the total negative magnetic flux as a function of time in Figure~\ref{f6} (lower right panel). The result shows that from 16:10 to 17:40~UT during which the event 5 occurred, magnetic flux decreases rapidly by an amount of \sm$10^{20}$ Mx. Assuming a depth of magnetic energy dissipation $d=1000$~km and the mean flux density in the calculation box $B=500$~G, the magnetic energy loss $E= B F d/4\pi$ is estimated to be in the order of $5\times10^{29}$ ergs. The flux cancellation across the circular PIL could be closely related to the triggering and progress of this flare related to the 3D null-point reconnection. Using a topological model, \citet{fletcher01b} demonstrated that such a flux cancellation may cause rapid coronal field restructuring by reconnection at the null, which produces heating and subsequent material ejection from multiple locations along the footprint of the fan in the lower atmosphere. Our event showing a dynamic nature of brightenings and jetting activities may also fit into this picture. However, the QSL reconnection should still play a central role in explaining the sequential brightening along the circular ribbon.

\section{SUMMARY} \label{summary}
\begin{figure}
\epsscale{.9}
\plotone{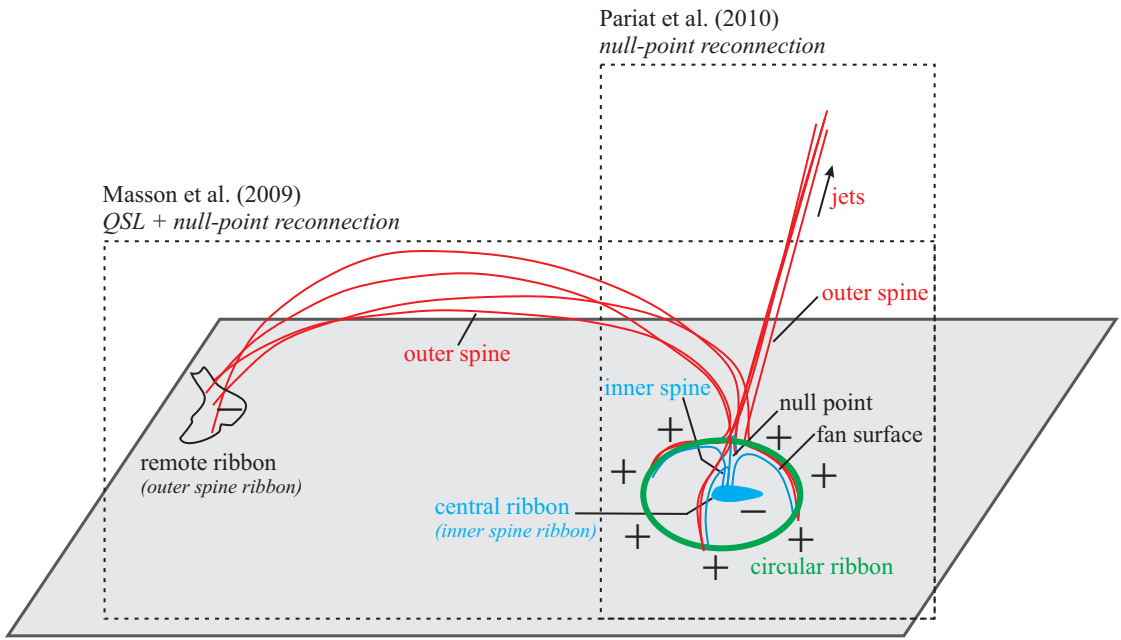}
\caption{Schematic picture demonstrating the relationship among circular flare ribbons, jets, and remote brightenings in a 3D null-point magnetic topology, based on and combining the modeling results of \citetalias{masson09} and \citetalias{pariat10}. See Section~\ref{introduction} for details. \label{f1}}
\end{figure}

In this paper, we have taken advantage of the high spatio-temporal resolution BBSO film observations that have recently been digitized to study five major flares characterized with circular chromospheric ribbons and surges. In particular, four flares originated from the same active region exhibit homologous properties and three of them are accompanied by remote brightenings. Circular ribbon flares were rarely reported before but are an important observational signature predicted as a result of the 3D null-point reconnection in the fan-spine magnetic topology. The high-quality \ha\ blue wing images allow us to study the event evolution in great detail in relation to the different types of 3D reconnection as proposed by \citetalias{masson09} and \citetalias{pariat10}. It is remarkable that different from the confined C-class event in \citetalias{masson09}, our ejective M- and X-class flares are clearly associated with surges; moreover, the model of \citetalias{pariat10} produces homologous jets, but does not accommodate remote brightenings as observed in our events (see Table~\ref{table1}). Therefore, the present observations point to a comprehensive picture combining the related topological magnetic field structure of \citetalias{masson09} and \citetalias{pariat10} (Figure~\ref{f1}). Our results can be summarized as follows.

\begin{enumerate}

\item The main attention is focused on the four homologous flares and surges on 1991 March 17--18 (events 1--4). By tracking the ribbon motion, we derive the magnetic flux change rates in the opposite polarities $\dot\phi_{+-}$, which are temporally correlated with the non-thermal flare emissions but are unbalanced. A caveat is that our method is only applicable for events with a 2D-like configuration. In fact, the existence of a 3D structure comprising the coronal null point and the related fan-spine magnetic structure are inferred from the unambiguous circular shape of the outer ribbon and the potential field extrapolation \citepalias{masson09}, as well as the drifting surge \citepalias{pariat10}.

\item In the event 1, the well-observed sequential brightening of the fan ribbon k2 along the circle and the elongated shape of the inner spine ribbon k1 strongly indicate the presence of extended QSLs embedding the separatrices and the slipping/slip-running reconnection \citepalias{masson09}. It is evidently observed that k1 undergoes a ``round-trip'' motion, presumably toward and then away from the coronal null point. As surges are produced through the null-point reconnection involving the outer spine \citepalias{pariat10}, a consistent observation could be that the surge begins the abrupt ejection after k1 reaches the turning point. We suggest that the intriguing motion of k1 may be due to the succession of slipping reconnection, first toward the null and then after the null reconnection, away from the null \citepalias{masson09}.

\item Remote brightenings are observed in the events 2--4, and are co-temporal with a second phase of flare non-thermal emissions about \sm1~minute after the main energy release. The delay might be related to the time of field slipping toward the null \citep{reid12}. The fact that the surge ejection is after the maximum of the remote brightening could indicate a quasi-instantaneous change of magnetic topology. These observations are the first signature that may indicate the opening and closing of the outer spine in the corona during a flare.

\item The successive occurrence of slipping/slip-run reconnection and null-point reconnection proposed by \citetalias{masson09} and \citet{reid12} is also manifested as two well separated phases of circular ribbon flare and ejective surge as observed in the event 5. The flaring source region experiences flux cancellation across the circular PIL, which may also contribute to the triggering of the null-point reconnection \citep{fletcher01}.

\end{enumerate}

It is important to understand how the 3D null-point system may be perturbed to be subjective to the subsequent reconnection by analyzing the magnetic field evolution \citepalias{masson09,pariat10}. This aspect requiring long-term magnetic field observation is out of the scope of this study. While the paucity of circular ribbon flares was suggested to be due to the possible uncommon presence of a single null point or the pronounced asymmetry of the fan \citepalias{masson09}, we emphasize that our survey in progress using the rich film data base has revealed a few more circular ribbon flares with the associated surges and/or remote brightenings. These events plus the continued advanced observations from the \Sdo\ have promise to shed further insights on the dynamics of magnetic reconnection in the null-point topology.

\acknowledgments

This study is dedicated to Professor Hal Zirin, the founder of Big Bear Solar Observatory, who passed away on 2012 January 3. We thank BBSO observing staffs for tremendous efforts in carrying out observations. We thank Jeff Nenow who developed nearly all the BBSO films and provided helpful information, John Varsik for help with VMG data, and the referee for highly detailed and valuable comments. The use of X-ray data from NOAA/NASA \goes\ project is gratefully acknowledged. C.L. is also grateful to Na Deng for helpful discussions. The work is supported by NSF under grants AGS 0839216 and AGS 0849453, and by NASA under grants NNX11AO70G and NNX11AC05G.

\newpage

\begin{deluxetable}{llcccccccc}
\tablecolumns{10}
\tablewidth{0pt}
\tablecaption{Events of Circular Ribbon Flares\label{table1}}
\tablehead{\colhead{} & \colhead{}  & \colhead{NOAA} & \colhead{Location} & \colhead{} & \colhead{\goes} & \colhead{Peak} & \colhead{Duration\tablenotemark{a}} & \colhead{Remote} & \colhead{} \\
\colhead{No} & \colhead{Date}  & \colhead{AR} & \colhead{(deg)} & \colhead{} & \colhead{Level} & \colhead{(UT)} & \colhead{(minutes)} & \colhead{brightening\tablenotemark{b}} & \colhead{Surge\tablenotemark{b}} }
\startdata
1 & 1991 Mar 17 & 6545 & W14, S10 & & M2 & 17:54 & 20 &  N & Y 	\\
2 & 1991 Mar 17 & 6545 & W16, S10 & & X1 & 21:29 & 10 &  Y & Y	\\
3 & 1991 Mar 18 & 6545 & W28, S11 & & M5 & 17:33 & 12 &  Y & Y 	\\
4 & 1991 Mar 18 & 6545 & W30, S11 & & M1 & 21:41 & 13 &  Y & N 	\\
5 & 1991 Feb 25 & 6509 & W22, S15 & & M1 & 17:15 & 32 &  N & Y   \\
\cutinhead{Related Observations/Models}
\citetalias{masson09} & 2002 Nov 16  & 10191 & S20, W21 & & C8.6 & 13:59 & 17 & Y & N \\
\citetalias{pariat10} & \nodata & \nodata & \nodata & & \nodata & \nodata & \nodata & N & Y \\

\enddata
\tablenotetext{a}{The duration is measured based on the time profile of \goes\ 1--8~\AA\ flux.}
\tablenotetext{b}{Y/N denote that the feature is present/absent in a specific event.}
\end{deluxetable}

\clearpage

\end{document}